\documentclass[prb,aps,amsmath,amssymb,showpacs,twocolumn]{revtex4-1}
\usepackage{amssymb,graphicx,float,dcolumn,bm,subfigure,color,colortbl}
\usepackage[dvipsnames]{xcolor}
\usepackage[titletoc,title]{appendix}
\usepackage[dotinlabels]{titletoc}
\usepackage[utf8]{inputenc}
\usepackage{nicefrac}
\usepackage{multirow, makecell, ragged2e}
\usepackage{cancel}
\usepackage{subfigure}
\usepackage{braket}
\usepackage{mathrsfs}
\usepackage[mathscr]{euscript}
\usepackage{hyperref}
\usepackage{ulem}

\newcommand{\be}{\begin{equation}}
\newcommand{\ee}{\end{equation}}

\newcommand{\ba}{\begin{eqnarray}}
\newcommand{\ea}{\end{eqnarray}}

\begin{document}

\title{Theoretical phase diagram of two-component composite fermions in double layer graphene}
\author{W. N. Faugno$^{1,2}$,  Ajit C. Balram$^{3}$, A. W\'ojs$^{4}$, and J. K. Jain$^1$}
\affiliation{$^1$Department of Physics, 104 Davey Lab, Pennsylvania State University, University Park, Pennsylvania 16802, USA}
\affiliation{$^2$Institut de Physique Th\'eorique, Universit\'e Paris-Saclay, CNRS, CEA, 91190 Gif sur Yvette, France}
\affiliation{$^{3}$The Institute of Mathematical Sciences, HBNI, CIT Campus, Chennai 600113, India}
\affiliation{$^{4}$Department of Theoretical Physics, Wroc\l{}aw University of Science and Technology, 50-370 Wroc\l{}aw, Poland}

\date{\today}

\begin{abstract} 
Theory predicts that double layer systems realize ``two-component composite fermions," which are formed when electrons capture both intra- and inter-layer vortices, to produce a wide variety of new strongly correlated liquid and crystal states as a function of the layer separation. Recent experiments in double layer graphene have revealed a large number of layer-correlated fractional quantum Hall states in the lowest Landau level, many of which have not been studied quantitatively in previous theoretical works. We consider the competition between various liquid and crystal states at several of these filling factors  (specifically, the states at total filling factors $\nu=3/7$, $4/9$, $6/11$, $4/7$, $3/5$, $2/3$, and $4/5$) to determine the theoretical phase diagram as a function of the layer separation. We compare our results with experiments and identify various observed states. In particular, we show that at small layer separations the states at total fillings $\nu=3/7$ and $\nu=3/5$ are partially pseudospin polarized, where pseudospin refers to the layer index. For certain fractions, such as $\nu=3/7$, interlayer correlations are predicted to survive to surprisingly large interlayer separations. 
\end{abstract}
\maketitle

\section{introduction}

The fractional quantum Hall effect (FQHE) in double layer systems is an interesting platform for studying strongly correlated electron systems. 
In single layers, there is essentially only one relevant energy scale for the FQHE. In a double layer system, two energy scales are present, namely the intralayer and interlayer Coulomb interactions. The competition between these two energy scales can be controlled by tuning the distance between the layers $d/l$, given here in units of magnetic length $l = \sqrt{\hbar/(eB)}$, where $B$ is the perpendicular magnetic field. As a result, double layer FQHE systems can realize many states beyond single layer FQHE systems. (A note on terminology: The term bilayer has been used for two uncoupled layers in the FQHE literature. However, in the graphene literature bilayer graphene refers to the system in which electrons can hop from one layer to another. We consider in this article systems where two graphene monolayers are in close proximity to one another, but separated by an insulating layer that suppresses electron tunneling. To avoid confusion, we use ``double layer" graphene for these systems.)

Very early on, Halperin generalized~\cite{Halperin83} the Laughlin states~\cite{Laughlin83} to multicomponent systems. In particular, a FQHE was predicted at total filling factor $\nu=1/2$, which was later observed experimentally~\cite{Suen92,Eisenstein92}. A much richer phase diagram for double layer FQHE was proposed by the composite fermion (CF) theory through construction of two-component Jain states~\cite{Scarola01b}, a generalization of the single component Jain states~\cite{Jain89}. The Halperin and Jain two-component states were considered in several theoretical articles~\cite{Chakraborty87, Yoshioka89, He91, He93, Park98, Scarola01b, Thiebaut14, Balram15, Balram15a}. In a closely related development, two-component states were considered in a single layer as well, where the two components are the electron spin~\cite{Park98,Jain07,Balram15a}. Double layer systems of fully spin polarized electrons in the limit $d/l=0$ are formally equivalent to a single layer system of spinful electrons with zero Zeeman energy, but the double layer physics is very different for nonzero $d/l$. For small $d/l$, the analogy to spinful composite fermions in a single layer is useful, and we will sometimes refer to the layer index as ``pseudospin" (with the two layers representing up and down pseudospins). The real spin will be taken as fully polarized, and thus will not be explicitly considered.

Recent experimental work in double layer graphene~\cite{Liu19,Li19} has given tremendous impetus to this topic, through the observation of a large number of FQH states, which brings the richness of double layer FQHE to the same level as FQHE in single layer systems~\cite{Csathy19}. Double layer graphene can explore parameter regimes not available to double layer systems in semiconductor quantum wells. In particular, these can access much smaller interlayer separations, as small as $d/l \sim 0.3$; in contrast, in semiconductor quantum wells it is difficult to attain values of $d/l$ less than 1 (because the separation must be greater than quantum well widths). Furthermore, in graphene double layers the separation $d/l$ can be varied continuously over a wide range by tuning the electron density. Conductance measurements have also been performed in the corbino geometry, probing the bulk properties of the sample and allowing for finer resolution of FQH states~\cite{Li19}. 

Our work in this article is motivated by the fact that while the experimental observations in Refs.~\cite{Liu19,Li19} are generally consistent with the theoretical predictions, the experiments observe layer-correlated states at many filling factors, such as $\nu=3/7$, 4/9, 6/11, 4/7, 3/5, 2/3, and 4/5, that were not studied quantitatively in Ref.~\cite{Scarola01b}. (The symbol $\nu$ denotes the {\it total} filling factor in this article.) The reason was that, as we shall see below, some of these states are described in terms of reverse-vortex attachment. While reverse-vortex attached states were studied quite early on~\cite{Wu93}, theoretical tools necessary to project states with reverse vortex attachment to the lowest Landau level (LLL) for large system sizes from which a reliable thermodynamic extrapolation can be made~\cite{Moller05,Davenport12} were only developed after the work of Ref.~\cite{Scarola01b}. Given the new experiments, we determine the phase diagram of FQHE at these filling factors as a function of the layer separation, to allow an identification of the experimentally observed states, and to make precise predictions that can be tested in future experimental work in double layer systems. Certain noteworthy features of our study are as follows. We consider {\it partially} pseudospin polarized states at $\nu = 3/7$ and 3/5, which are identified with the observed incompressible states at these filling factors. We consider additional candidate states at $\nu=1/2$, but do not find them to be energetically favorable. We find that the interlayer correlations are strongly filling factor dependent, and for certain filling factors they extend to much larger interlayer separations than previously thought. The principal result of our present study is the phase diagram is presented below in Fig.~\ref{fig:PhaseDiagram}. 

Several mysteries still remain. Fractions 10/17, 10/13, and 6/7 are unexplained by our calculation here as there is no simple correlated two-component Jain state for these fillings (certain complicated constructions for pseudospin singlet states at these fillings are not considered here). Also, at $\nu=6/7$ and $\nu=4/5$, drag experiments indicate a single interlayer vortex, which is expected for $\nu=4/5$, but surprising at $\nu =6/7$ (see the discussion section for some candidate states). These two fractions have been interpreted in terms of an interlayer pairing of composite fermions~\cite{Liu19}. The microscopic origin of such pairing is presently unclear. 

The plan of our paper is as follows. In Sec.~\ref{sec:candidate_states} we present the various two-component candidate states considered in this work. Next, we discuss in Sec.~\label{sec:methods} the methods of variational Monte Carlo and exact diagonalization which have been used to evaluate the energies of the various candidate states. We conclude the paper in Sec.~\ref{sec:results} with a discussion of our results.

\section{Candidate States}
\label{sec:candidate_states}
We present the trial wave functions we will be considering in the subsections below. There are two main types of states considered: two-component Jain states and two-component electron / CF crystal states. We also consider more exotic states for filling factor $\nu=1/2$ that go beyond the standard CF theory. It is expected that for small $d/l$, the most favorable states will be those with definite pseudospin. For large $d/l$, on the other hand, one expects two uncorrelated layers. In the intermediate region, in general, several two-component Jain states or two-component crystals are possible, with diminishing inter-layer correlations as $d/l$ increases. 

To simplify the discussion, we present all wave functions below in the disk geometry. These can be translated in the standard manner~\cite{Jain07} into the spherical geometry, which is used in all of our calculations below.

\begin{table*}
\label{tab:States}
\begin{center}
\begin{tabular} { | c | c | c | c | c|}

\hline
$\nu$ & Pseudospin eigenstate & $m=2$ & $m=1$ & $m=0$ \\
\hline
1/3 & $\Phi_1^3(\{z_\uparrow\},\{z_\downarrow\})$ & {$(1/4,\ 1/4|\ 2)$} & {$(1/5,\ 1/5|\ 1)$} & {$(1/6,\ 1/6|\ 0)$}\\
\hline
2/5 & $\Phi_{1,1}\Phi_1^2$ &---& {$(1/4,\ 1/4|\ 1)$} & {$(1/5,\ 1/5|\ 0)$}\\
\hline
3/7 & $\mathcal{P}_{\rm LLL}\Phi_{2,1}\Phi_1^{2}$ & $(3/8,\ 3/8|\ 2)^{\dagger}$ & {$(3/11,\ 3/11|\ 1)$} & {$(3/14,\ 3/14|\ 0)$}\\
\hline
4/9 & $\mathcal{P}_{\rm LLL}\Phi_{2,2}\Phi_1^2$ & --- & {$(2/7,\ 2/7|\ 1)$} & {$(2/9,\ 2/9|\ 0)$}\\
\hline
6/11 & $\mathcal{P}_{\rm LLL}\Phi_{3,3}\Phi_1^2$ & --- & $(3/8,\ 3/8|\ 1)^{\dagger}$ & {$(3/11,\ 3/11|\ 0)$}\\
\hline
4/7 & $\mathcal{P}_{\rm LLL}\Phi_{2,2}^{*}\Phi_1^2$ & --- & {$(2/5,\ 2/5|\ 1)$} & {$(2/7,\ 2/7|\ 0)$}\\
\hline
10/17 & --- & $(5/7,\ 5/7|\ 2)^{\dagger}$ & $(5/12,\ 5/12|\ 1)^{\dagger}$ & $(5/17,\ 5/17|\ 0)^{\dagger}$\\
\hline
3/5 & $\mathcal{P}_{\rm LLL}\Phi_{2,1}^{*}\Phi_1^2$ & $(3/4,\ 3/4|\ 2)^{\dagger}$ & {$(3/7,\ 3/7|\ 1)$} & {$(3/10,\ 3/10|\ 0)$}\\
\hline
2/3 & $\mathcal{P}_{\rm LLL}\Phi_{1,1}^{*} \Phi_1^2$ & $(1,\ 1|\ 2)^{\dagger}$ & {$(1/2,\ 1/2|\ 1)$} & {$(1/3,\ 1/3|\ 0)$}\\
\hline
3/4 & --- & $(3/2,\ 3/2|\ 2)^{\dagger}$ & {$(3/5,\ 3/5|\ 1)$} & $(3/8,\ 3/8|\ 0)^{\dagger}$\\
\hline
10/13 & --- & $(5/3,\ 5/3|\ 2)^{\dagger}$ & $(5/8,\ 5/8|\ 1)^{\dagger}$ & $(5/13,\ 5/13|\ 0)^{\dagger}$\\
\hline
4/5 & --- & $(2,\ 2|\ 2)^{\dagger}$ & {$(2/3,\ 2/3|\ 1)$} & {$(2/5,\ 2/5|\ 0)$}\\
\hline
6/7 & --- & $(3,\ 3|\ 2)^{\dagger}$ & $(3/4,\ 3/4|\ 1)^{\dagger}$ & {$(3/7,\ 3/7|\ 0)$}\\
\hline
\end{tabular}
\end{center}
\caption{Candidate ground state wave functions. We list all wave functions for each filling factor observed, sans $1/2$, which we list in a subsequent section. Where possible we use the notation $(\bar{\nu},\ \bar{\nu}|\ m)$, where $m$ is the number of interlayer vortices attached and $\bar{\nu}$ is the effective filling of each layer. For pseudospin states that we cannot write in this form we write the wave function as $\Phi_{n^\uparrow,n^\downarrow}\Phi_1^{2p}$ with $\Phi_{n^\uparrow,n^\downarrow}=\Phi_{n^\uparrow}\Phi_{n^\downarrow}$. $\Phi_n$ is the IQHE Slater determinant at filling factor $n$. The $^{\dagger}$ sign marks states that are not amenable to Monte Carlo calculations since they cannot be projected to the lowest Landau level using the Jain-Kamilla method (see text for details). For entries marked with a --- we have not been able to identify suitable candidate ground state wave functions and appropriate interlayer correlation strengths.}
\end{table*}

\subsection{States in the limit of small $d/l$}
In the limit of zero layer separation, the Coulomb interaction is pseudospin independent and the system becomes equivalent to a single layer system of spinful electrons with zero Zeeman energy. The states obtained in the limit that $d/l\rightarrow 0$ have been well studied both theoretically and experimentally~\cite{Zhang84,Yoshioka86,Xie89,Wu93,Du95,Park98,Yeh99,Kukushkin99,Park01,Balram15a,Zhang16}. The wave function can thus be written in terms of pseudospins where $\uparrow$ labels one layer and $\downarrow$ labels the other.
The full wave function at total filling factor $\nu$ is given by 
\begin{equation}
\mathcal{A}\left[\Psi_\nu(\{z_j^\uparrow\},\{z_j^\downarrow\})\alpha_1...\alpha_{N_\uparrow}\beta_1...\beta_{N_\downarrow} \right]
\label{eq:generic_two_component}
\end{equation}
where $\mathcal{A}$ is the antisymmetrization operator, $\Psi_\nu$ is the spatial part of the wave function  at filling $\nu$, $\alpha$ and $\beta$ are spinors corresponding to layer pseudospins, $N_\uparrow$ and $N_\downarrow$ are the number of electrons with pseudospin $\uparrow$ and $\downarrow$ respectively, $N=N_\uparrow+N_\downarrow$ is the total number of electrons and $z^\alpha_{j}=x^\alpha_{j}-iy^\alpha_{j}$ is the two-dimensional coordinate of the electron parametrized as a complex number with $\alpha$ being the pseudospin of the particle. Because the full wave function must have a definite pseudospin, the spatial part $\Psi_\nu(\{z_j^\uparrow\},\{z_j^\downarrow\})$ must satisfy Fock's cyclic condition~\cite{Hamermesh62}. 
We specialize to the $\nu=n/(2pn\pm 1)$, which maps into $n$ filled Landau levels of composite fermions [termed Lambda levels ($\Lambda$Ls)]. Here, in general, we have $n=n_{\uparrow}+n_{\downarrow}$, where $n_{\uparrow}$ and $n_{\downarrow}$ are the filling factors of the up and down pseudospin $\Lambda$Ls. In this case, we have 
\begin{equation}
\Psi_\nu(\{z_j\})=\mathcal{P}_{\text{LLL}}\Phi_{n_{\uparrow}}(\{z^{\uparrow}\})\Phi_{n_{\downarrow}}(\{z^{\downarrow}\})\prod_{1\leq j<k\leq N}(z_j-z_k)^{2p}
\label{eq:two_component_Jain_CF},
\end{equation}
where $\Phi_n$ is the wave function of IQH state with $n$ filled LLs,  
and $\mathcal{P}_{\text{LLL}}$ is the LLL projection operator. We evaluate the projection using the standard Jain-Kamila (JK) projection method~\cite{Jain97}. The product in the Jastrow factor $\Phi_1=\prod_{i<j}(z_i-z_j)$ extends over all pairs of particles, independent of their pseudospin. 

The above wave function corresponds to the highest weight state with $S=S_z=(N^{\uparrow}-N^{\downarrow})/2$. We will assume in this work that the densities in the two layers are equal, i.e. $N^{\uparrow}=N^{\downarrow}=N/2$ and therefore $S_z=0$. For pseudospin singlet states we already have $S_z=0$. For other cases, the $S_z=0$ state can in principle be constructed by repeated application of the pseudospin lowering operator $S^-$. For fully pseudospin polarized states i.e. when the total pseudospin $S = N/2$, the spatial portion of the wave function remains the same upon action of the $S^-$ operator because the spatial portion is fully antisymmetric. In such cases, the spatial part of the $S_z=0$ state is obtained simply by assigning half of the coordinates to one layer and the rest to the other.  Partially pseudospin polarized states do not allow for such a simple construction. The action of the $S^-$ operator on the highest weight state produces a sum of many different Slater determinants, which quickly becomes intractable for use as trial wave functions for variational Monte Carlo (VMC) calculations. We address transitions at filling factors $\nu=3/5$ and $3/7$, where partially polarized states arise, by using exact diagonalization.

 \subsection{Two-component Jain states}
 
 For $d/l \neq 0$, the ground state wave function no longer has to have a definite total pseudospin, and many other candidate states become possible. We consider two-component states that are labeled $(\bar{\nu},\ \bar{\nu}|\ m)$, whose wave functions are given by 
 \begin{equation}
 \Psi_{(\bar{\nu},\ \bar{\nu}|\ m)}=\Psi_{\bar{\nu}}(\{z^\uparrow_i\})\Psi_{\bar{\nu}}(\{z^\downarrow_i\})\prod_{i,j}(z^\uparrow_i-z^\downarrow_j)^{m},
 \label{two-component-w.f.}
 \end{equation}
 where we have assumed equal densities in the two layers. 
 These states are constructed by placing each layer in an independent FQH state at an effective filling $\bar{\nu}$ and allowing for additional attachment of $m$ vortices between electrons in different layers. In general, the effective filling in the $\uparrow$ layer, $\bar{\nu_\uparrow}$, is defined in terms of the number of particles in each layer and the number of flux quanta, $N_\phi$, as $\bar{\nu_\uparrow}=N_\uparrow/(N_\phi - mN_\downarrow)$. For systems with equal electron densities in the two layers, the effective filling factor $\bar{\nu}$ is related to the total filling $\nu$ by $\nu=2\bar{\nu}/(1+m\bar{\nu})$ or $\bar{\nu} = \nu/(2-m\nu)$. 
When $\bar{\nu} = n/(2pn\pm1)$, we have $\Psi_{\bar{\nu}} = \mathcal{P}_\text{LLL} \Phi_n\Phi_1^{2p}$ and $\nu=2n/[(2p+m)n\pm 1]$. The $\left({n\over 2pn\pm1}, {n\over 2pn\pm1} |m  \right)$ state describes an incompressible state where composite fermions with $2p$ intralayer vortices and $m$ interlayer vortices attached to them undergo $\nu^*=n$ integer quantum Hall effect (IQHE). The Halperin states are obtained when $n=1$. 
 
\subsection{Two-component crystal states}
In addition to incompressible states, two component electron or CF crystals have also been predicted in double layers at filling factors $\nu = 1/3$ , 1/5, and 2/5~\cite{Faugno18}. The wave functions for these crystals are given by
\begin{equation}
\Psi_{\nu}^{\text{X}(2p,m)} = \Psi_{\bar{\nu}}^{\text{X}(2p)} (\{z_i^\uparrow\})
\Psi_{\bar{\nu}}^{\text{X}(2p)} (\{z_i^\downarrow\})
\prod_{i,j}(z_i^\uparrow-z_j^\downarrow)^m
\end{equation}
with 
\begin{equation}
\Psi_{\bar{\nu}}^{\text{X}(2p)}(\{z_i\}) = \text{Det}[\phi_{R_i}(z_j)] \prod_{j<k}(z_j-z_k)^{2p},
\end{equation}
\begin{equation}
\phi_{R_i}(z) = \frac{1}{\sqrt{2\pi}}\exp{\left[\frac{1}{2}\bar{R_i}z-\frac{1}{4}|z|^2-\frac{1}{4}|R_i|^2\right]}
\end{equation}
Here $\phi_{R_i}(z)$ is an electron wave packet localized at position $R_i$;  $\text{Det}[\phi_{R_i}(z_j)]$ is the wave function of a single layer crystal of electrons located at positions $\{R_i \}$; $\Psi_{\bar{\nu}}^{\text{X}(2p)}(\{z_i\}) $ is the wave function of a crystal of composite fermions in a single layer; and $\Psi_{\nu}^{\text{X}(2p,m)}$ is the wave function of a layer-correlated double layer crystal. The superscript X denotes the crystal type. Previous work by Faugno {\it et al.}~\cite{Faugno18} has shown that the most likely crystal structures in bilayer systems are triangular Ising antiferromagnetic (TIAF), correlated square (CS) and binary graphene (BG). 

\subsection{Exotic candidates at $\nu=1/2$}

At $\nu=1/2$, we consider several additional candidate states beyond those listed above. We note that it has been previously shown that two-component crystals are not relevant for this system~\cite{Faugno18}. For $\nu = 1/2$, 
we consider compressible states, namely fully pseudospin polarized CF Fermi sea (CFFS)
\begin{equation}
\Psi_{\text{FS}(1/2,\ 1/2|\ 2)}= \mathcal{P}_{\rm LLL}\Phi_{\text{FS}}(\{z_i\})\prod_{1\leq i<j\leq N}(z_i-z_j)^2
\end{equation}
and the pseudospin singlet CFFS
\begin{equation}
\Psi_{(1/2_\text{FS},\ 1/2_\text{FS}| 2)}=\mathcal{P}_{\rm LLL}\Phi_{\text{FS}}(\{z_i^\uparrow\})\Phi_{\text{FS}}(\{z_i^\downarrow\})\prod_{1\leq i<j\leq N}(z_i-z_j)^2
\end{equation}
where $\Phi_{\text{FS}}$ is the wave function for the Fermi sea and the Jastrow factor involves all coordinates.
In addition we consider the $(1/4,\ 1/4|\ 0)$ state:
 \begin{eqnarray}
 \Psi_{(1/4,\ 1/4|\ 0)} &=& \left[\mathcal{P}_\text{LLL}\Phi_{\text{FS}}(\{z_i^\uparrow\})\prod_{1\leq j<k=N/2} (z_i^\uparrow-z_j^\uparrow)^4\right] \nonumber\\
 & \times & \left[\mathcal{P}_\text{LLL}\Phi_{\text{FS}}(\{z_i^\downarrow\})\prod_{1\leq j<k=N/2} (z_i^\downarrow-z_j^\downarrow)^4\right]~~~ 
 \end{eqnarray}
The incompressible $(1/3,\ 1/3|\ 1)$ state is equivalent to the Halperin 331 state,

We can further construct a set of non-Abelian states by substituting the Fermi sea wave function with the Pfaffian wave function ${\rm Pf}[(z_i-z_j)^{-1}]$~\cite{Moore91}. The Pfaffian of an antisymmetric matrix with elements $M_{i,j}$ is given by $\mathcal{A}(M_{1,2}M_{3,4}...M_{N-1,N})$. This allows us to construct several additional wave functions as shown in Table~\ref{tab:12states}. These states are interesting because they support Majorana modes at the vortices and the edge of the system.

Finally, we consider states constructed via the parton theory~\cite{Jain89b}. In the parton theory of the FQHE, each electron is broken into $m$ parts, called partons, each with a fractional charge $-\nu e/\nu_i$ where $\nu$ is the total filling for electrons, $\nu_i$ is the filling for each parton species and $(-e)$ is the charge of the electron. We obtain an incompressible state when all $\nu_i=n_i$ for $n_i$ integers. Each state is labeled by  $n_1n_2n_3...$ and to denote negative fillings we use $\bar{n}_i = -n_i$. The wave functions suggested by this theory are then products over a series of IQH Slater determinants, $\mathcal{P}_{\rm LLL}\prod_i\Phi_{n_i}$. At $\nu = 1/2$, the most likely parton states are 221, $\bar{2}\bar{2}111$, and a two-component $1,121$ state. The 221 state~\cite{Wu16,Bandyopadhyay18,Kim18} is not considered here as it cannot be projected to the LLL by JK projection. On the other hand the $\bar{2}\bar{2}111$ state can be projected as
\begin{equation}
\Psi_{\bar{2}\bar{2}111}=\mathcal{P}_\text{LLL}\Phi_{2}^*\Phi_{2}^*\Phi_1^3 \sim \Psi_{2/3}^2/\Phi_1
\end{equation}
where the $\sim$ sign indicates that states either side of the sign differ in the details of the projection. We do not expect such details to affect the topological properties of the state~\cite{Balram16b}. The $\bar{2}\bar{2}111$ state is in the same topological class as the anti-Pfaffian~\cite{Levin07,Lee07}, but has been shown to be a better candidate at $\nu=5/2$ than the traditional anti-Pfaffian~\cite{Balram18}. The two component $1,1211$ state can also be evaluated via JK projection as
\begin{equation}
\Psi_{1,121} = \Phi_{1,1}\Phi_2\Phi_1 \sim \Phi_{1,1}\Psi_{2/5}/\Phi_1
\end{equation}
where $\Phi_{1,1}$ is the product of two $n=1$ IQH Slater determinants each containing a distinct set of half the total number of particles.

\begin{table}
\begin{center}
\begin{tabular}{| c | c |}
\hline
$(1/2_\text{FS},\ 1/2_\text{FS}| 2)$ & $\mathcal{P}_{\rm LLL}\Phi_{\text{FS}}(\{z_i^\uparrow\})\Phi_{\text{FS}}(\{z_i^\downarrow\})\Phi_1^2$\\
\hline
$\text{FS}(1/2,\ 1/2|\ 2)$ & $\mathcal{P}_{\rm LLL}\Phi_{\text{FS}}\Phi_1^2$\\
\hline
$(1/2_\text{PF},\ 1/2_\text{PF}| 2)$ & $\text{Pf}(\frac{1}{z_i^\uparrow-z_j^\uparrow})\text{Pf}(\frac{1}{z_i^\downarrow-z_j^\downarrow})\Phi_1^2(\{z_i\})$\\
\hline
$\text{PF}(1/2,\ 1/2|\ 2)$ & $\text{Pf}(\frac{1}{z_i-z_j})\Phi_1^2(\{z_i\})$\\
\hline
$\bar{2}\bar{2}111$ & $\mathcal{P}_{\rm LLL}\Phi_2^*\Phi_2^*\Phi_1^3$\\
\hline
$221$ & $\mathcal{P}_{\rm LLL}\Phi_2\Phi_2\Phi_1$\\
\hline
$1,121$ & $\mathcal{P}_{\rm LLL}\Phi_{1,1}\Phi_2\Phi_1$\\
\hline
$(1/3,\ 1/3|\ 1)$ & $\Phi_1^3(\{z_i^\uparrow\})\Phi_1^3(\{z_i^\downarrow\})\prod_{i,j}(z_i^\uparrow-z_j^\downarrow)$\\
\hline
$(1/4_\text{FS},\ 1/4_\text{FS}|\ 0)$ & $\Phi_{\text{FS}}(\{z_i^\uparrow\})\Phi_1^4(\{z_i^\uparrow\})\Phi_{\text{FS}}(\{z_i^\downarrow\})\Phi_1^4(\{z_i^\downarrow\})$\\
\hline
$(1/4_\text{PF},\ 1/4_\text{PF}|\ 0)$ & $\text{Pf}(\frac{1}{z_i^\uparrow-z_j^\uparrow})\text{Pf}(\frac{1}{z_i^\downarrow-z_j^\downarrow})\Phi_1^4(\{z_i^\uparrow\})\Phi_1^4(\{z_i^\downarrow\})$\\
\hline
\end{tabular}
\caption{Candidate wave functions at $\nu = 1/2$. We consider states from the theory of two-component Jain construction as well as more exotic non-Abelian Pfaffian states and states constructed from the parton theory. The method for constructing each of the proposed wave functions is discussed in detail in the text. The only states relevant for the parameters considered in our work are the singlet CFFS, the Halperin 331 state, and the uncoupled two-component $(1/4\ 1/4|\ 0)$.}
\label{tab:12states}
\end{center}
\end{table}

\section{Methods of Calculation}
\label{sec:methods}
In this work, we use the spherical geometry~\cite{Haldane83} wherein $N$ electrons are placed on a sphere with a magnetic monopole of strength $2Q$ placed at its center. The radius of the sphere is $\sqrt{Q}l$. It is convenient to define spinor coordinates $u$ and $v$, which are related to the spherical polar and azimuthal angles $\theta$ and $\phi$ by $u = \cos(\theta/2)e^{i\phi/2}$ and $v = \sin(\theta/2)e^{-i\phi/2}$. The chord distance, in units of $l$, between particles $i$ and $j$ on the sphere is $r_{i,j}=2\sqrt{Q}|u_iv_j-u_jv_i|$, and the Jastrow factor is given by $\prod_{i<j}(u_iv_j-u_jv_i)$. To compare the various candidate ground state wave functions we evaluate their double layer Coulomb interaction which is given by
\begin{eqnarray}
V_{\uparrow,\uparrow} = V_{\downarrow,\downarrow} = \frac{e^2}{\epsilon l}\frac{1}{r_{i,j}}\\
V_{\uparrow,\downarrow} = \frac{e^2}{\epsilon l} \frac{1}{\sqrt{r_{i,j}^2+(d/l)^2}}
\label{eqn:interaction}
\end{eqnarray}
where  $\epsilon$ is the dielectric constant of the material.

Determining the best candidate wave function at a given value of $d/l$ is a question of energetics. We evaluate the energy of each state under the interaction of Eq.~(\ref{eqn:interaction}) via Monte-Carlo integration. We achieve an error less than $10^{-5}$ $e^2/\epsilon l$ for a Monte-Carlo simulation with $10^7$ iterations. We additionally multiply the energies by $\sqrt{2Q\nu/N}$, the ratio of the density in the finite system to the density in the thermodynamic limit, to suppress the dependence of the energy on the particle number~\cite{Morf86}.

We constructed our candidate states for many different finite system sizes up to 100 particles for states with parallel vortex attachment and 40 particles for states with reverse vortex attachment. JK projection of reverse vortex attached states is carried out using the scheme of Ref.~\cite{Davenport12} which requires computationally expensive high precision arithmetic. We then carry out a linear fitting of the energies as a function of $1/N$ to determine the energy in the thermodynamic limit. 
Because the interactions involving the background can be complicated in double layer systems, we extrapolate the energy difference between states, choosing one base state to compare the rest with for each filling factor. (We note that since all states cannot necessarily be constructed at all system sizes, we interpolate the electron-electron interaction of the base state before taking the difference.) Thermodynamic extrapolations for various states at each filling factor for a separation of $d/l =1$ are shown in the panels of Fig.~\ref{Fig:thermo}. We make several approximations in our calculation. We assume there is no Landau level mixing, the electron spin is frozen by the magnetic field, and disorder is negligible. We also assume no tunneling between layers, which is achieved experimentally in double layer graphene by including an insulating layer of hexagonal boron nitride between the graphene layers.

To study the partially polarized states at $\nu = 3/7$ and $\nu = 3/5$, we find the ground state of the LLL Coulomb interaction for a spinful electron system at zero Zeeman energy and in the spin sector with $S_z=0$ for even $N$ and $S_{z}=1/2$ for odd $N$ using exact diagonalization (ED). We take this state as a good representation of the partially polarized CF states wherein CFs occupy two spin up and one spin down $\Lambda$Ls with parallel vortex attachment for 3/7 and reverse vortex attachment for 3/5. The expectation value of the energy of the state is then evaluated as a function of layer separation under the interaction given in Eq.~(\ref{eqn:interaction}) for system sizes 5, 8, and 11 for 3/7 and 5, 8, 11, and 14 for 3/5. We obtain the interaction energy, including the electron-electron and electron-background and background-background contributions, in the thermodynamic limit. For this purpose, we correct for the finite size deviation of the density from its thermodynamic value by multiplying the total energy by $\sqrt{2Q\nu/N}$~\cite{Morf86}. The background-background and electron-background interactions are given by $V_{bb} + V_{eb} = -N^2/4\sqrt{Q}-N^2/4\sqrt{Q+(d/l)^2}$. The extrapolations for $\nu=3/7$ and $\nu=3/5$ are shown in Fig.~\ref{Fig:ppthermo}. In order to compare with the VMC results, we obtain the thermodynamic energy of the reference state at each filling factor including electron-electron and electron-background and background-background contributions.

\section{Results and Discussion}
\label{sec:results}

\begin{figure*}
\includegraphics[width=7in]{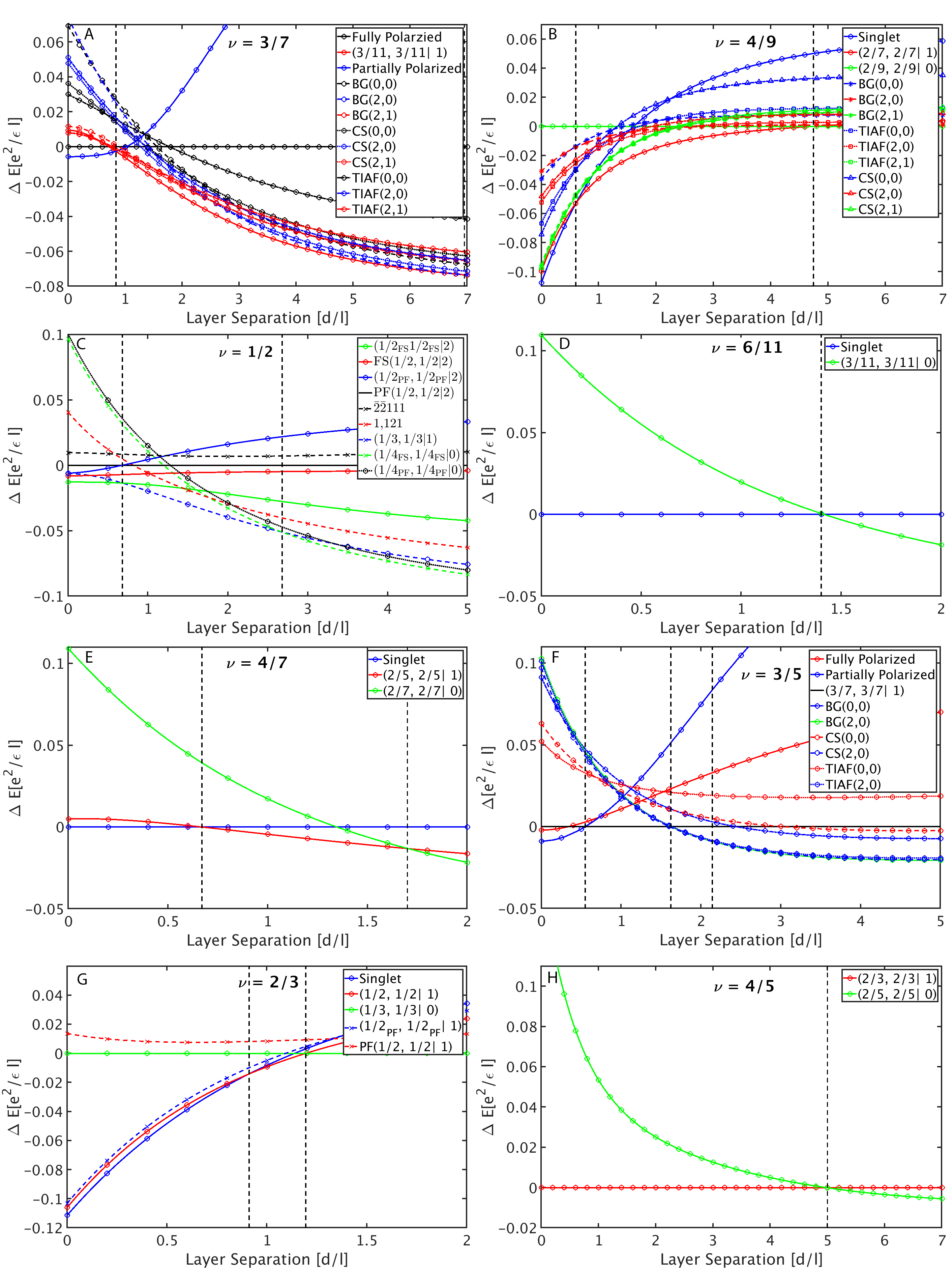}
\caption{Energies for various candidate states as a function of layer separation $d/l$. All energies are measured relative to a convenient reference state. The liquid states are labeled by $(\bar{\nu},\ \bar{\nu}|\ m)$, the same labeling as in \ref{tab:States} and \ref{tab:12states}. TIAF$(2p,m)$, CS$(2p,m)$ and BG$(2p,m)$ correspond to CF crystals with structures of triangular Ising antiferromagnetic (TIAF), correlated square (CS) and binary graphene (BG) with $2p$ the number of intralayer vortices and $m$ the number of interlayer vortices.}\label{Fig:Evd}
\end{figure*}

The thermodynamic extrapolations of the energies of various candidate states are shown in the Appendix~\ref{app:thermodynamic_limits} for a typical interlayer separation of $d/l=1$. The thermodynamic energies obtained in a similar fashion in a range of $d/l$ are shown in Fig.~\ref{Fig:Evd}. At each filling factor, we choose one state as a convenient reference state, and all energies are measured relative to the energy of this reference state. The meanings of the labels for different candidate states are explained in the text. The labels ``singlet," ``fully polarized," and ``partially polarized" refer to pseudospin (i.e. the layer index). The $d/l$ values where transitions occur are indicated by the vertical dashed lines.

\begin{figure*}
\includegraphics[width=7in]{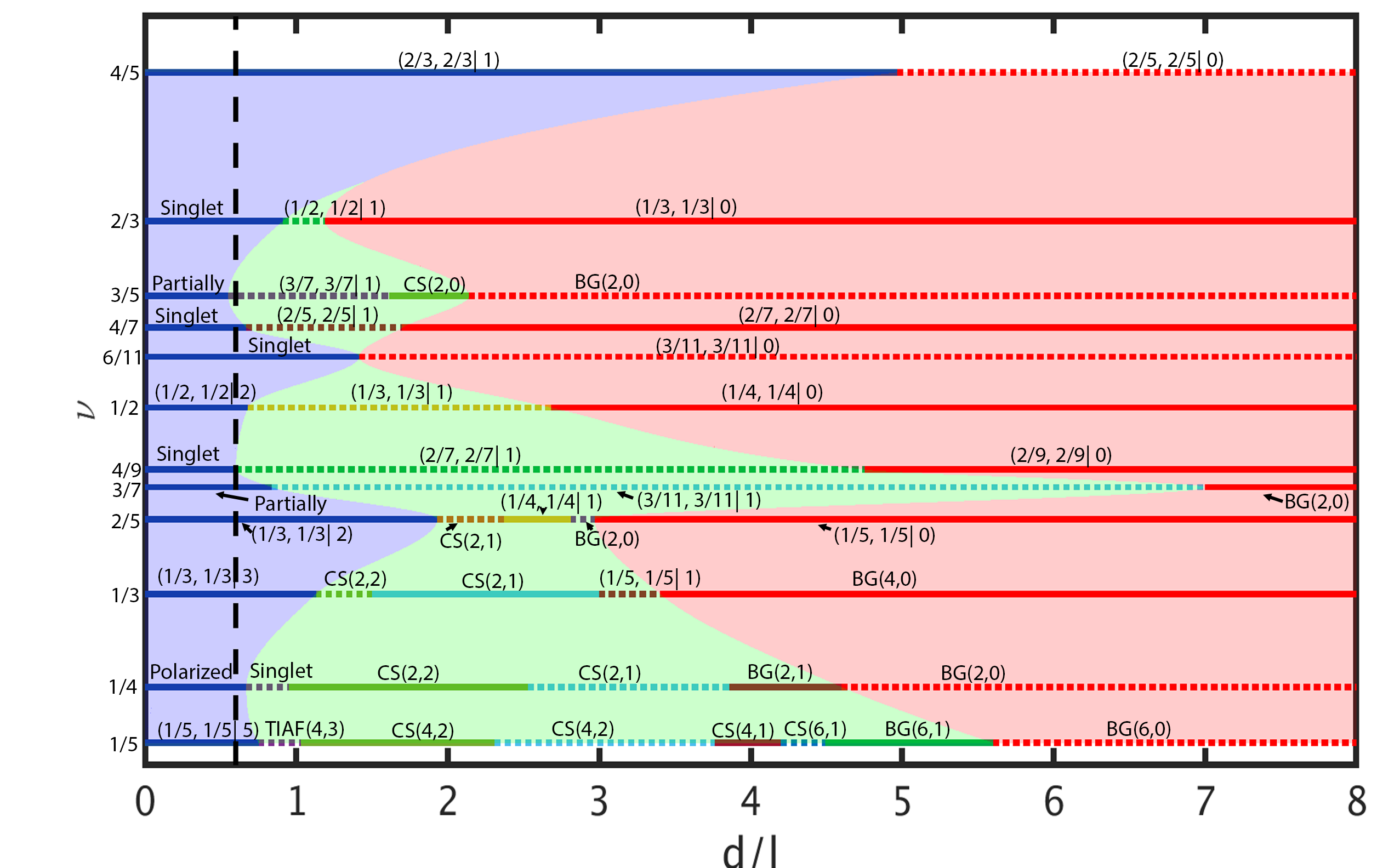}
\caption{Theoretical phase diagram of double layer states. The labbel $(\bar{\nu},~\bar{\nu}|~  m)$ refers to the state given in Eq.~(\ref{two-component-w.f.}), where $\bar{\nu}$ the effective single layer filling factor and $m$ the number of interlayer vortices. Pseudospin singlet and partially polarized states are labeled as such directly. TIAF$(2p,m)$, CS$(2p,m)$ and BG$(2p,m)$ correspond to CF crystals with structures of triangular Ising antiferromagnetic (TIAF), correlated square (CS) and binary graphene (BG) with $2p$ the number of intralayer vortices and $m$ the number of interlayer vortices. The region corresponding to the $d/l = 0$ limit is colored in blue, whereas the region where the states are layer-uncorrelated is colored in red. The vertical dashed black line marks the upper limit of the separations where the experimental measurements have been reported~\cite{Li19,Liu19}. For completeness, we have included results from previous studies; the results for filling factors 1/5, 1/3, and 2/5 are taken from \cite{Scarola01b} and \cite{Faugno18} and for filling factors 1/4 from \cite{Faugno19}. } 
\label{fig:PhaseDiagram}
\end{figure*}

The phase diagram as a function of the layer separation and the filling factor, obtained from the calculations shown in Fig.~\ref{Fig:Evd}, is shown in Fig.~\ref{fig:PhaseDiagram}. This phase diagram is the principal result of our study. (For completeness, it includes results from previous studies~\cite{Scarola01b,Faugno18,Faugno19}.) The vertical dashed line markes the largest $d/l$ in the experiments of Refs.~\cite{Liu19,Li19}, which shows that the experiments lie within the $d/l = 0$ limit of our phase diagram. The overall trend shows that as the layer separation is increased the interlayer correlations weaken in favor of stronger intralayer correlations, eventually producing states at large $d/l$ that do not have interlayer correlations. This is consistent with previous theoretical calculations that focused on filling factors below $\nu = 1/2$ \cite{Scarola01b,Faugno18}. Despite this overall trend, we find that the nature of the states and the strength of their interlayer correlations are strongly filling factor dependent, with the onset of the layer-uncorrelated regime ranging from 1.2$l$ at $\nu=2/3$ to 7$l$ at $\nu=3/7$.

In the limit of $d/l = 0$, the system is equivalent to spinful electrons at zero Zeeman energy, and the ground states should be the same with spin replaced by pseudospin. As a result, the ground states for $\nu = 3/7$ and $3/5$ are partially pseudospin polarized states with a minimal value of $|S_z|$, the pseudospin operator that counts the difference between the number of particles in each layer. We find that these states persist for finite values of $d/l$. These states have not been considered previously in the context of pseudospin. A clear experimental signature of such states is that they should survive for finite density imbalances between layers, corresponding to increasing the layer polarization until $|S_z|$ achieves its maximal value.

One phenomenon of note is the persistence of interlayer correlations up to large $d/l$ at fillings $\nu = 1/5$, $3/7$, $4/9$, and $4/5$. The layer correlated states are favored up to $d/l = 5$ at $\nu=3/5$, $3/7$, and $1/5$, and even up to $d/l = 7$ at $\nu=3/7$. This is in stark contrast to other fillings, for example those studied in Ref.~\cite{Scarola01b}, where the layer-uncorrelated state occurs beyond a layer separation of $d/l = 1.5-3$. The robustness of certain states is also surprising. For example, even though the $3/11$ state in a single layer is a fragile FQHE state, the $(3/11,\ 3/11 |\ 1)$ state at $\nu=3/7$ appears rather robust. Similar feature had been found in the phase digram of spinful composite fermions in a single layer~\cite{Balram15}.

Our phase diagram presents previous results at filling factors $\nu = 1/5$, 1/4, 1/3, 2/5, and 1/2~\cite{Scarola01b,Faugno18,Faugno19}. For filling factors 1/5, 1/4, 1/3, and 2/5, we see strong competition between two-component Jain states and CF crystals in the intermediate separation regime. We note that at $\nu=1/4$ theory does not predict any incompressible ground states, but there is a pseudospin phase transition from a fully polarized pseudospin state to a pseudospin singlet~\cite{Faugno19}. At $\nu = 1/2$, we find that only two-component Jain states are relevant.

At $\nu = 2/3$, our calculation suggests the possibility of an intermediate state consisting of two coupled CF Fermi seas, denoted as $(1/2,\ 1/2|\ 1)$. Previous ED calculations find a direct transition from the layer singlet 2/3 to the layer uncorrelated $(1/3,\ 1/3|\ 0)$~\cite{Bonesteel96, McDonald96}. ED calculations, however, are not able to deal with compressible states in a reliable manner due to finite size limitations. 

We have only considered in this work states of the type $(\bar{\nu},\ \bar{\nu}|\ m)$ where $\bar{\nu}$ belong to the primary Jain sequence for non-interacting composite fermions, i.e. $\bar{\nu}=n/(2pn\pm1)$. Further, we allow for reverse vortex attachment only within each layer but not between layers.
Even within this class, many states are not amenable to our VMC calculations with sufficiently large systems, and are therefore not considered. We give here some examples. At $\nu = 4/5$ and 6/7, there are candidate states for spinful composite fermions in a single layer, constructed from combinations of particle-hole conjugation and reverse vortex attachment. These states can be written using the notation of Ref.~\cite{Balram15} as $\overline{[[1,1]_{-2}]}_{-2}$ at $\nu=4/5$ and $\overline{[\overline{[[1,1]_{-2}]}_{-2}]}_{-2}$ at $\nu=6/7$~\cite{Balram16c}. Similarly, either a fully spin polarized or a partially polarized candidate state at $\nu = 10/13$ can be constructed from parent states at $\nu^{*}=10/7=1+3/7$ by reverse vortex attachment~\cite{Balram15}. We have not considered the double layer analogs of these states in this work.  At $\nu= 10/13$ and 10/17, we can construct double layer incompressible FQH states $(5/3,\ 5/3|\ 2)$ and $(5/7,\ 5/7|\ 2)$. These have not been considered above because of technical reasons.

Of course, it is also possible to consider real spin in addition to the layer pseudospin. That enlarges the space to SU(4), which allows for new states beyond those constructed here~\cite{Toke07,Wu15}.

We mention several unexplained observations.  The Hall plateau at $\nu=1/2$, commonly associated with the Halperin 331 state, persists to lower values of $d/l$ than we predict. The nature of the observed state at $\nu=6/7$ 
is not well understood in a quantitative sense. 
Coulomb drag experiments show a single interlayer zero at this filling factor, but the state $(3/4,\ 3/4|\ 1)$ is not incompressible for non-interacting composite fermions. 
It has been suggested that the state at $\nu=6/7$ arises due to pairing of composite fermions~\cite{Li19, Liu19}. As stated above, we are not able to calculate the phase diagram at several filling factors, such as $\nu= 10/13$ and 10/17.

In summary, double layer graphene systems have made it possible to study two component FQHE states in a larger parameter regime than before. 
That has motivated us to evaluate the theoretical phase diagram including many states not previously considered, revealing the richness of states available in these systems. The current experimental data \cite{Li19,Liu19} appear to lie more or less in the $d/l=0$ limit of the phase diagram, but the rest of the theoretical phase diagram should be experimentally accessible in graphene based systems. These systems thus provide an ideal platform for furthering our understanding of strongly correlated electron systems and the competition between the inter and intra layer correlations.

\begin{acknowledgments}
The work at Penn State (W.N.F. and J.K.J.) was supported by the U. S. Department of Energy under Grant no. DE-SC0005042. W.N.F. thanks the Chateaubriand Fellowship, and Thierry Jolicoeur for discussions and hospitality at CNRS. This project has received funding from the Polish NCN Grant No. 2014/14/A/ST3/00654 (A. W.). We thank Wroc\l{}aw Centre for Networking and Supercomputing and Academic Computer Centre CYFRONET, both parts of PL-Grid Infrastructure. Some portions of this research were conducted with Advanced CyberInfrastructure computational resources provided by The Institute for CyberScience at The Pennsylvania State University. 
\end{acknowledgments}

\begin{appendix}

\section{Thermodynamic limits}
\label{app:thermodynamic_limits}

In this appendix, we show the thermodynamic extrapolations of the energies of various candidate states for $d/l=1$. Similar extrapolations at other values of $d/l$ are used to deduce the phase digram shown in the main text.

\begin{figure*}
\includegraphics[width=7in]{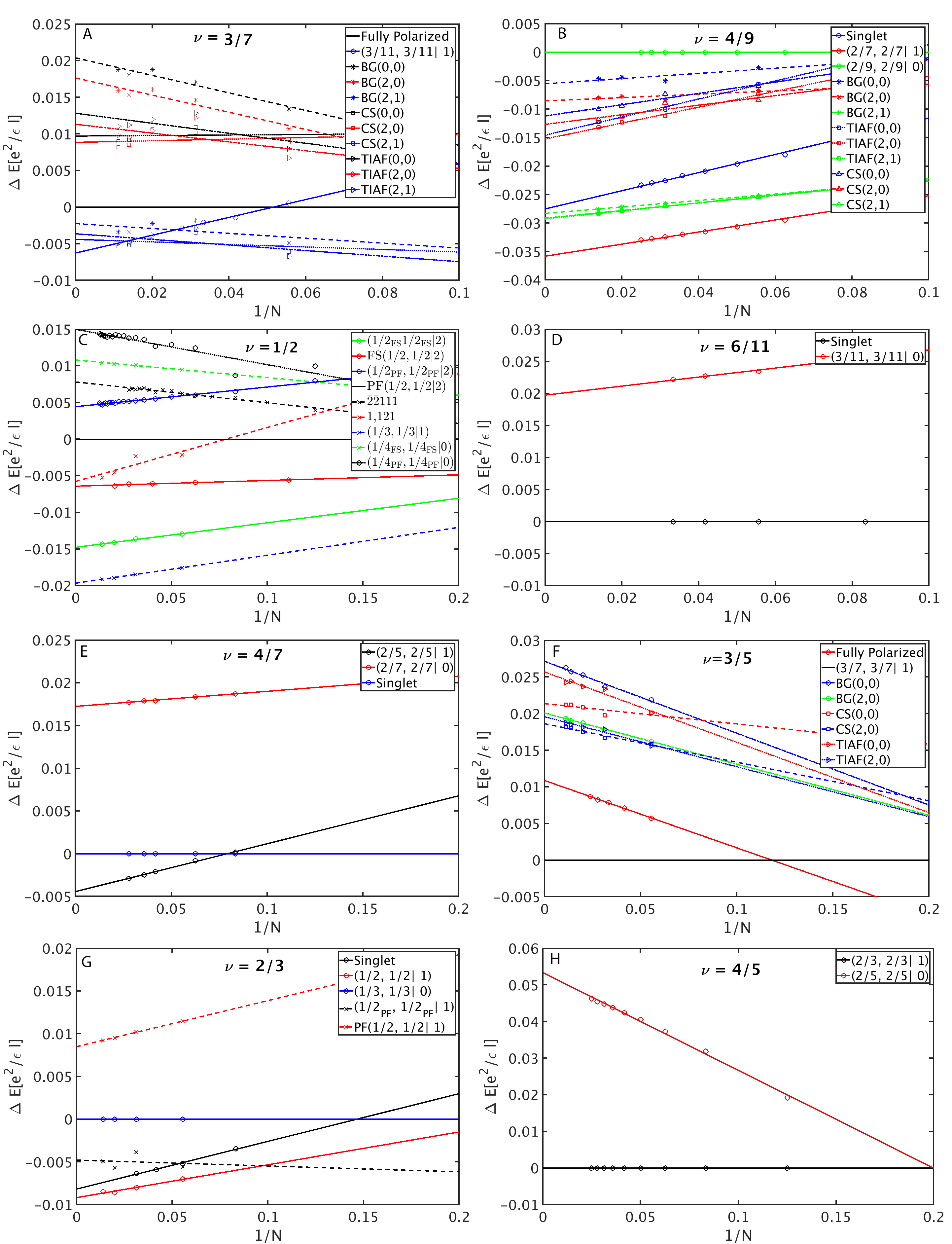}
\caption{Thermodynamic limits of the energies of various candidate states for $d/l = 1$. All energies are measured relative to a convenient reference state. The liquid states are labeled by $(\bar{\nu},\ \bar{\nu}|\ m)$, the same labeling as in \ref{tab:States} and \ref{tab:12states}. TIAF$(2p,m)$, CS$(2p,m)$ and BG$(2p,m)$ correspond to CF crystals with structures of triangular Ising antiferromagnetic (TIAF), correlated square (CS) and binary graphene (BG) with $2p$ the number of intralayer vortices and $m$ the number of interlayer vortices.}\label{Fig:thermo}
\end{figure*}

\begin{figure*}
\includegraphics[width=7in]{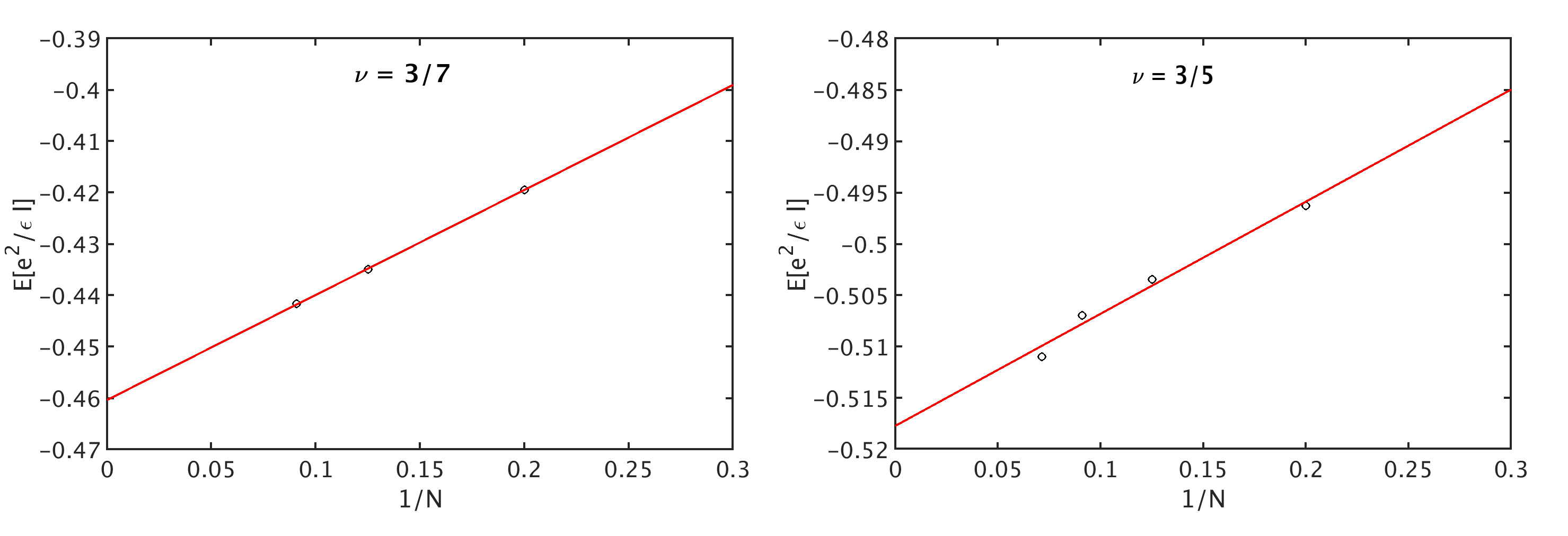}
\caption{Thermodynamic limits of the energies of the partially polarized trial states at $\nu=3/7$ and $\nu=3/5$. The trial wave function is generated by exact diagonalization at $d/l=0$. Its Coulomb energy is then calculated as a function of $d/l$. This figure shows results for $d/l = 1$. This energy includes the background-background and electron-background contribution, $V_{bb}+V_{eb}=-N^2/4\sqrt{Q}-N^2/4\sqrt{Q+(d/l)^2}$.}\label{Fig:ppthermo}
\end{figure*}

\end{appendix}


\begin{thebibliography}{49}
\expandafter\ifx\csname natexlab\endcsname\relax\def\natexlab#1{#1}\fi
\expandafter\ifx\csname bibnamefont\endcsname\relax
  \def\bibnamefont#1{#1}\fi
\expandafter\ifx\csname bibfnamefont\endcsname\relax
  \def\bibfnamefont#1{#1}\fi
\expandafter\ifx\csname citenamefont\endcsname\relax
  \def\citenamefont#1{#1}\fi
\expandafter\ifx\csname url\endcsname\relax
  \def\url#1{\texttt{#1}}\fi
\expandafter\ifx\csname urlprefix\endcsname\relax\def\urlprefix{URL }\fi
\providecommand{\bibinfo}[2]{#2}
\providecommand{\eprint}[2][]{\url{#2}}

\bibitem[{\citenamefont{Halperin}(1983)}]{Halperin83}
\bibinfo{author}{\bibfnamefont{B.~I.} \bibnamefont{Halperin}},
  \bibinfo{journal}{Helvetica Physica Acta} \textbf{\bibinfo{volume}{56}},
  \bibinfo{pages}{75} (\bibinfo{year}{1983}), ISSN \bibinfo{issn}{0018-0238}.

\bibitem[{\citenamefont{Laughlin}(1983)}]{Laughlin83}
\bibinfo{author}{\bibfnamefont{R.~B.} \bibnamefont{Laughlin}},
  \bibinfo{journal}{Phys. Rev. Lett.} \textbf{\bibinfo{volume}{50}},
  \bibinfo{pages}{1395} (\bibinfo{year}{1983}),
  \urlprefix\url{http://link.aps.org/doi/10.1103/PhysRevLett.50.1395}.

\bibitem[{\citenamefont{Suen et~al.}(1992)\citenamefont{Suen, Engel, Santos,
  Shayegan, and Tsui}}]{Suen92}
\bibinfo{author}{\bibfnamefont{Y.~W.} \bibnamefont{Suen}},
  \bibinfo{author}{\bibfnamefont{L.~W.} \bibnamefont{Engel}},
  \bibinfo{author}{\bibfnamefont{M.~B.} \bibnamefont{Santos}},
  \bibinfo{author}{\bibfnamefont{M.}~\bibnamefont{Shayegan}}, \bibnamefont{and}
  \bibinfo{author}{\bibfnamefont{D.~C.} \bibnamefont{Tsui}},
  \bibinfo{journal}{Phys. Rev. Lett.} \textbf{\bibinfo{volume}{68}},
  \bibinfo{pages}{1379} (\bibinfo{year}{1992}),
  \urlprefix\url{http://link.aps.org/doi/10.1103/PhysRevLett.68.1379}.

\bibitem[{\citenamefont{Eisenstein et~al.}(1992)\citenamefont{Eisenstein,
  Boebinger, Pfeiffer, West, and He}}]{Eisenstein92}
\bibinfo{author}{\bibfnamefont{J.~P.} \bibnamefont{Eisenstein}},
  \bibinfo{author}{\bibfnamefont{G.~S.} \bibnamefont{Boebinger}},
  \bibinfo{author}{\bibfnamefont{L.~N.} \bibnamefont{Pfeiffer}},
  \bibinfo{author}{\bibfnamefont{K.~W.} \bibnamefont{West}}, \bibnamefont{and}
  \bibinfo{author}{\bibfnamefont{S.}~\bibnamefont{He}}, \bibinfo{journal}{Phys.
  Rev. Lett.} \textbf{\bibinfo{volume}{68}}, \bibinfo{pages}{1383}
  (\bibinfo{year}{1992}),
  \urlprefix\url{http://link.aps.org/doi/10.1103/PhysRevLett.68.1383}.

\bibitem[{\citenamefont{Scarola and Jain}(2001)}]{Scarola01b}
\bibinfo{author}{\bibfnamefont{V.~W.} \bibnamefont{Scarola}} \bibnamefont{and}
  \bibinfo{author}{\bibfnamefont{J.~K.} \bibnamefont{Jain}},
  \bibinfo{journal}{Phys. Rev. B} \textbf{\bibinfo{volume}{64}},
  \bibinfo{pages}{085313} (\bibinfo{year}{2001}),
  \urlprefix\url{http://link.aps.org/doi/10.1103/PhysRevB.64.085313}.

\bibitem[{\citenamefont{Jain}(1989{\natexlab{a}})}]{Jain89}
\bibinfo{author}{\bibfnamefont{J.~K.} \bibnamefont{Jain}},
  \bibinfo{journal}{Phys. Rev. Lett.} \textbf{\bibinfo{volume}{63}},
  \bibinfo{pages}{199} (\bibinfo{year}{1989}{\natexlab{a}}),
  \urlprefix\url{http://link.aps.org/doi/10.1103/PhysRevLett.63.199}.

\bibitem[{\citenamefont{Chakraborty and Pietil\"ainen}(1987)}]{Chakraborty87}
\bibinfo{author}{\bibfnamefont{T.}~\bibnamefont{Chakraborty}} \bibnamefont{and}
  \bibinfo{author}{\bibfnamefont{P.}~\bibnamefont{Pietil\"ainen}},
  \bibinfo{journal}{Phys. Rev. Lett.} \textbf{\bibinfo{volume}{59}},
  \bibinfo{pages}{2784} (\bibinfo{year}{1987}),
  \urlprefix\url{https://link.aps.org/doi/10.1103/PhysRevLett.59.2784}.

\bibitem[{\citenamefont{Yoshioka et~al.}(1989)\citenamefont{Yoshioka,
  MacDonald, and Girvin}}]{Yoshioka89}
\bibinfo{author}{\bibfnamefont{D.}~\bibnamefont{Yoshioka}},
  \bibinfo{author}{\bibfnamefont{A.~H.} \bibnamefont{MacDonald}},
  \bibnamefont{and} \bibinfo{author}{\bibfnamefont{S.~M.}
  \bibnamefont{Girvin}}, \bibinfo{journal}{Phys. Rev. B}
  \textbf{\bibinfo{volume}{39}}, \bibinfo{pages}{1932} (\bibinfo{year}{1989}),
  \urlprefix\url{https://link.aps.org/doi/10.1103/PhysRevB.39.1932}.

\bibitem[{\citenamefont{He et~al.}(1991)\citenamefont{He, Xie, Das~Sarma, and
  Zhang}}]{He91}
\bibinfo{author}{\bibfnamefont{S.}~\bibnamefont{He}},
  \bibinfo{author}{\bibfnamefont{X.~C.} \bibnamefont{Xie}},
  \bibinfo{author}{\bibfnamefont{S.}~\bibnamefont{Das~Sarma}},
  \bibnamefont{and} \bibinfo{author}{\bibfnamefont{F.~C.} \bibnamefont{Zhang}},
  \bibinfo{journal}{Phys. Rev. B} \textbf{\bibinfo{volume}{43}},
  \bibinfo{pages}{9339} (\bibinfo{year}{1991}),
  \urlprefix\url{https://link.aps.org/doi/10.1103/PhysRevB.43.9339}.

\bibitem[{\citenamefont{He et~al.}(1993)\citenamefont{He, Das~Sarma, and
  Xie}}]{He93}
\bibinfo{author}{\bibfnamefont{S.}~\bibnamefont{He}},
  \bibinfo{author}{\bibfnamefont{S.}~\bibnamefont{Das~Sarma}},
  \bibnamefont{and} \bibinfo{author}{\bibfnamefont{X.~C.} \bibnamefont{Xie}},
  \bibinfo{journal}{Phys. Rev. B} \textbf{\bibinfo{volume}{47}},
  \bibinfo{pages}{4394} (\bibinfo{year}{1993}),
  \urlprefix\url{http://link.aps.org/doi/10.1103/PhysRevB.47.4394}.

\bibitem[{\citenamefont{Park and Jain}(1998)}]{Park98}
\bibinfo{author}{\bibfnamefont{K.}~\bibnamefont{Park}} \bibnamefont{and}
  \bibinfo{author}{\bibfnamefont{J.~K.} \bibnamefont{Jain}},
  \bibinfo{journal}{Phys. Rev. Lett.} \textbf{\bibinfo{volume}{80}},
  \bibinfo{pages}{4237} (\bibinfo{year}{1998}),
  \urlprefix\url{http://link.aps.org/doi/10.1103/PhysRevLett.80.4237}.

\bibitem[{\citenamefont{Thiebaut et~al.}(2014)\citenamefont{Thiebaut, Goerbig,
  and Regnault}}]{Thiebaut14}
\bibinfo{author}{\bibfnamefont{N.}~\bibnamefont{Thiebaut}},
  \bibinfo{author}{\bibfnamefont{M.~O.} \bibnamefont{Goerbig}},
  \bibnamefont{and} \bibinfo{author}{\bibfnamefont{N.}~\bibnamefont{Regnault}},
  \bibinfo{journal}{Phys. Rev. B} \textbf{\bibinfo{volume}{89}},
  \bibinfo{pages}{195421} (\bibinfo{year}{2014}),
  \urlprefix\url{https://link.aps.org/doi/10.1103/PhysRevB.89.195421}.

\bibitem[{\citenamefont{Balram et~al.}(2015{\natexlab{a}})\citenamefont{Balram,
  T\"oke, W\'ojs, and Jain}}]{Balram15}
\bibinfo{author}{\bibfnamefont{A.~C.} \bibnamefont{Balram}},
  \bibinfo{author}{\bibfnamefont{C.}~\bibnamefont{T\"oke}},
  \bibinfo{author}{\bibfnamefont{A.}~\bibnamefont{W\'ojs}}, \bibnamefont{and}
  \bibinfo{author}{\bibfnamefont{J.~K.} \bibnamefont{Jain}},
  \bibinfo{journal}{Phys. Rev. B} \textbf{\bibinfo{volume}{91}},
  \bibinfo{pages}{045109} (\bibinfo{year}{2015}{\natexlab{a}}),
  \urlprefix\url{http://link.aps.org/doi/10.1103/PhysRevB.91.045109}.

\bibitem[{\citenamefont{Balram et~al.}(2015{\natexlab{b}})\citenamefont{Balram,
  T\"oke, W\'ojs, and Jain}}]{Balram15a}
\bibinfo{author}{\bibfnamefont{A.~C.} \bibnamefont{Balram}},
  \bibinfo{author}{\bibfnamefont{C.}~\bibnamefont{T\"oke}},
  \bibinfo{author}{\bibfnamefont{A.}~\bibnamefont{W\'ojs}}, \bibnamefont{and}
  \bibinfo{author}{\bibfnamefont{J.~K.} \bibnamefont{Jain}},
  \bibinfo{journal}{Phys. Rev. B} \textbf{\bibinfo{volume}{92}},
  \bibinfo{pages}{075410} (\bibinfo{year}{2015}{\natexlab{b}}),
  \urlprefix\url{http://link.aps.org/doi/10.1103/PhysRevB.92.075410}.

\bibitem[{\citenamefont{Jain}(2007)}]{Jain07}
\bibinfo{author}{\bibfnamefont{J.~K.} \bibnamefont{Jain}},
  \emph{\bibinfo{title}{Composite Fermions}} (\bibinfo{publisher}{Cambridge
  University Press, New York, US}, \bibinfo{year}{2007}).

\bibitem[{\citenamefont{Liu et~al.}(2019)\citenamefont{Liu, Hao, Watanabe,
  Taniguchi, Halperin, and Kim}}]{Liu19}
\bibinfo{author}{\bibfnamefont{X.}~\bibnamefont{Liu}},
  \bibinfo{author}{\bibfnamefont{Z.}~\bibnamefont{Hao}},
  \bibinfo{author}{\bibfnamefont{K.}~\bibnamefont{Watanabe}},
  \bibinfo{author}{\bibfnamefont{T.}~\bibnamefont{Taniguchi}},
  \bibinfo{author}{\bibfnamefont{B.~I.} \bibnamefont{Halperin}},
  \bibnamefont{and} \bibinfo{author}{\bibfnamefont{P.}~\bibnamefont{Kim}},
  \bibinfo{journal}{Nature Physics} \textbf{\bibinfo{volume}{15}},
  \bibinfo{pages}{893} (\bibinfo{year}{2019}), ISSN \bibinfo{issn}{1745-2481},
  \urlprefix\url{https://doi.org/10.1038/s41567-019-0546-0}.

\bibitem[{\citenamefont{Li et~al.}(2019)\citenamefont{Li, Shi, Zeng, Watanabe,
  Taniguchi, Hone, and Dean}}]{Li19}
\bibinfo{author}{\bibfnamefont{J.~I.~A.} \bibnamefont{Li}},
  \bibinfo{author}{\bibfnamefont{Q.}~\bibnamefont{Shi}},
  \bibinfo{author}{\bibfnamefont{Y.}~\bibnamefont{Zeng}},
  \bibinfo{author}{\bibfnamefont{K.}~\bibnamefont{Watanabe}},
  \bibinfo{author}{\bibfnamefont{T.}~\bibnamefont{Taniguchi}},
  \bibinfo{author}{\bibfnamefont{J.}~\bibnamefont{Hone}}, \bibnamefont{and}
  \bibinfo{author}{\bibfnamefont{C.~R.} \bibnamefont{Dean}},
  \bibinfo{journal}{Nature Physics} \textbf{\bibinfo{volume}{15}},
  \bibinfo{pages}{898} (\bibinfo{year}{2019}), ISSN \bibinfo{issn}{1745-2481},
  \urlprefix\url{https://doi.org/10.1038/s41567-019-0547-z}.

\bibitem[{\citenamefont{Csathy and Jain}(2019)}]{Csathy19}
\bibinfo{author}{\bibfnamefont{G.~A.} \bibnamefont{Csathy}} \bibnamefont{and}
  \bibinfo{author}{\bibfnamefont{J.~K.} \bibnamefont{Jain}},
  \bibinfo{journal}{NATURE PHYSICS} \textbf{\bibinfo{volume}{15}},
  \bibinfo{pages}{883} (\bibinfo{year}{2019}), ISSN \bibinfo{issn}{1745-2473}.

\bibitem[{\citenamefont{Wu et~al.}(1993)\citenamefont{Wu, Dev, and
  Jain}}]{Wu93}
\bibinfo{author}{\bibfnamefont{X.~G.} \bibnamefont{Wu}},
  \bibinfo{author}{\bibfnamefont{G.}~\bibnamefont{Dev}}, \bibnamefont{and}
  \bibinfo{author}{\bibfnamefont{J.~K.} \bibnamefont{Jain}},
  \bibinfo{journal}{Phys. Rev. Lett.} \textbf{\bibinfo{volume}{71}},
  \bibinfo{pages}{153} (\bibinfo{year}{1993}),
  \urlprefix\url{http://link.aps.org/doi/10.1103/PhysRevLett.71.153}.

\bibitem[{\citenamefont{M\"oller and Simon}(2005)}]{Moller05}
\bibinfo{author}{\bibfnamefont{G.}~\bibnamefont{M\"oller}} \bibnamefont{and}
  \bibinfo{author}{\bibfnamefont{S.~H.} \bibnamefont{Simon}},
  \bibinfo{journal}{Phys. Rev. B} \textbf{\bibinfo{volume}{72}},
  \bibinfo{pages}{045344} (\bibinfo{year}{2005}),
  \urlprefix\url{http://link.aps.org/doi/10.1103/PhysRevB.72.045344}.

\bibitem[{\citenamefont{Davenport and Simon}(2012)}]{Davenport12}
\bibinfo{author}{\bibfnamefont{S.~C.} \bibnamefont{Davenport}}
  \bibnamefont{and} \bibinfo{author}{\bibfnamefont{S.~H.} \bibnamefont{Simon}},
  \bibinfo{journal}{Phys. Rev. B} \textbf{\bibinfo{volume}{85}},
  \bibinfo{pages}{245303} (\bibinfo{year}{2012}),
  \urlprefix\url{http://link.aps.org/doi/10.1103/PhysRevB.85.245303}.

\bibitem[{\citenamefont{Zhang and Chakraborty}(1984)}]{Zhang84}
\bibinfo{author}{\bibfnamefont{F.~C.} \bibnamefont{Zhang}} \bibnamefont{and}
  \bibinfo{author}{\bibfnamefont{T.}~\bibnamefont{Chakraborty}},
  \bibinfo{journal}{Phys. Rev. B} \textbf{\bibinfo{volume}{30}},
  \bibinfo{pages}{7320} (\bibinfo{year}{1984}),
  \urlprefix\url{https://link.aps.org/doi/10.1103/PhysRevB.30.7320}.

\bibitem[{\citenamefont{Yoshioka}(1986)}]{Yoshioka86}
\bibinfo{author}{\bibfnamefont{D.}~\bibnamefont{Yoshioka}},
  \bibinfo{journal}{Journal of the Physical Society of Japan}
  \textbf{\bibinfo{volume}{55}}, \bibinfo{pages}{885} (\bibinfo{year}{1986}),
  \eprint{http://dx.doi.org/10.1143/JPSJ.55.885},
  \urlprefix\url{http://dx.doi.org/10.1143/JPSJ.55.885}.

\bibitem[{\citenamefont{Xie et~al.}(1989)\citenamefont{Xie, Guo, and
  Zhang}}]{Xie89}
\bibinfo{author}{\bibfnamefont{X.~C.} \bibnamefont{Xie}},
  \bibinfo{author}{\bibfnamefont{Y.}~\bibnamefont{Guo}}, \bibnamefont{and}
  \bibinfo{author}{\bibfnamefont{F.~C.} \bibnamefont{Zhang}},
  \bibinfo{journal}{Phys. Rev. B} \textbf{\bibinfo{volume}{40}},
  \bibinfo{pages}{3487} (\bibinfo{year}{1989}),
  \urlprefix\url{http://link.aps.org/doi/10.1103/PhysRevB.40.3487}.

\bibitem[{\citenamefont{Du et~al.}(1995)\citenamefont{Du, Yeh, Stormer, Tsui,
  Pfeiffer, and West}}]{Du95}
\bibinfo{author}{\bibfnamefont{R.~R.} \bibnamefont{Du}},
  \bibinfo{author}{\bibfnamefont{A.~S.} \bibnamefont{Yeh}},
  \bibinfo{author}{\bibfnamefont{H.~L.} \bibnamefont{Stormer}},
  \bibinfo{author}{\bibfnamefont{D.~C.} \bibnamefont{Tsui}},
  \bibinfo{author}{\bibfnamefont{L.~N.} \bibnamefont{Pfeiffer}},
  \bibnamefont{and} \bibinfo{author}{\bibfnamefont{K.~W.} \bibnamefont{West}},
  \bibinfo{journal}{Phys. Rev. Lett.} \textbf{\bibinfo{volume}{75}},
  \bibinfo{pages}{3926} (\bibinfo{year}{1995}),
  \urlprefix\url{http://link.aps.org/doi/10.1103/PhysRevLett.75.3926}.

\bibitem[{\citenamefont{Yeh et~al.}(1999)\citenamefont{Yeh, Stormer, Tsui,
  Pfeiffer, Baldwin, and West}}]{Yeh99}
\bibinfo{author}{\bibfnamefont{A.~S.} \bibnamefont{Yeh}},
  \bibinfo{author}{\bibfnamefont{H.~L.} \bibnamefont{Stormer}},
  \bibinfo{author}{\bibfnamefont{D.~C.} \bibnamefont{Tsui}},
  \bibinfo{author}{\bibfnamefont{L.~N.} \bibnamefont{Pfeiffer}},
  \bibinfo{author}{\bibfnamefont{K.~W.} \bibnamefont{Baldwin}},
  \bibnamefont{and} \bibinfo{author}{\bibfnamefont{K.~W.} \bibnamefont{West}},
  \bibinfo{journal}{Phys. Rev. Lett.} \textbf{\bibinfo{volume}{82}},
  \bibinfo{pages}{592} (\bibinfo{year}{1999}),
  \urlprefix\url{http://link.aps.org/doi/10.1103/PhysRevLett.82.592}.

\bibitem[{\citenamefont{Kukushkin et~al.}(1999)\citenamefont{Kukushkin,
  v.~Klitzing, and Eberl}}]{Kukushkin99}
\bibinfo{author}{\bibfnamefont{I.~V.} \bibnamefont{Kukushkin}},
  \bibinfo{author}{\bibfnamefont{K.}~\bibnamefont{v.~Klitzing}},
  \bibnamefont{and} \bibinfo{author}{\bibfnamefont{K.}~\bibnamefont{Eberl}},
  \bibinfo{journal}{Phys. Rev. Lett.} \textbf{\bibinfo{volume}{82}},
  \bibinfo{pages}{3665} (\bibinfo{year}{1999}),
  \urlprefix\url{http://link.aps.org/doi/10.1103/PhysRevLett.82.3665}.

\bibitem[{\citenamefont{Park and Jain}(2001)}]{Park01}
\bibinfo{author}{\bibfnamefont{K.}~\bibnamefont{Park}} \bibnamefont{and}
  \bibinfo{author}{\bibfnamefont{J.}~\bibnamefont{Jain}},
  \bibinfo{journal}{Solid State Communications} \textbf{\bibinfo{volume}{119}},
  \bibinfo{pages}{291 } (\bibinfo{year}{2001}), ISSN \bibinfo{issn}{0038-1098},
  \urlprefix\url{http://www.sciencedirect.com/science/article/pii/S0038109801001053}.

\bibitem[{\citenamefont{Zhang et~al.}(2016)\citenamefont{Zhang, W\'ojs, and
  Jain}}]{Zhang16}
\bibinfo{author}{\bibfnamefont{Y.}~\bibnamefont{Zhang}},
  \bibinfo{author}{\bibfnamefont{A.}~\bibnamefont{W\'ojs}}, \bibnamefont{and}
  \bibinfo{author}{\bibfnamefont{J.~K.} \bibnamefont{Jain}},
  \bibinfo{journal}{Phys. Rev. Lett.} \textbf{\bibinfo{volume}{117}},
  \bibinfo{pages}{116803} (\bibinfo{year}{2016}),
  \urlprefix\url{http://link.aps.org/doi/10.1103/PhysRevLett.117.116803}.

\bibitem[{\citenamefont{Hamermesh}(1962)}]{Hamermesh62}
\bibinfo{author}{\bibfnamefont{M.}~\bibnamefont{Hamermesh}},
  \emph{\bibinfo{title}{Group Theory and Its Application to Physical Problems}}
  (\bibinfo{publisher}{New York: Dover, US}, \bibinfo{year}{1962}).

\bibitem[{\citenamefont{Jain and Kamilla}(1997)}]{Jain97}
\bibinfo{author}{\bibfnamefont{J.~K.} \bibnamefont{Jain}} \bibnamefont{and}
  \bibinfo{author}{\bibfnamefont{R.~K.} \bibnamefont{Kamilla}},
  \bibinfo{journal}{Int. J. Mod. Phys. B} \textbf{\bibinfo{volume}{11}},
  \bibinfo{pages}{2621} (\bibinfo{year}{1997}).

\bibitem[{\citenamefont{Faugno et~al.}(2018)\citenamefont{Faugno, Duthie,
  Wales, and Jain}}]{Faugno18}
\bibinfo{author}{\bibfnamefont{W.~N.} \bibnamefont{Faugno}},
  \bibinfo{author}{\bibfnamefont{A.~J.} \bibnamefont{Duthie}},
  \bibinfo{author}{\bibfnamefont{D.~J.} \bibnamefont{Wales}}, \bibnamefont{and}
  \bibinfo{author}{\bibfnamefont{J.~K.} \bibnamefont{Jain}},
  \bibinfo{journal}{Phys. Rev. B} \textbf{\bibinfo{volume}{97}},
  \bibinfo{pages}{245424} (\bibinfo{year}{2018}),
  \urlprefix\url{https://link.aps.org/doi/10.1103/PhysRevB.97.245424}.

\bibitem[{\citenamefont{Moore and Read}(1991)}]{Moore91}
\bibinfo{author}{\bibfnamefont{G.}~\bibnamefont{Moore}} \bibnamefont{and}
  \bibinfo{author}{\bibfnamefont{N.}~\bibnamefont{Read}},
  \bibinfo{journal}{Nucl. Phys. B} \textbf{\bibinfo{volume}{360}},
  \bibinfo{pages}{362 } (\bibinfo{year}{1991}), ISSN \bibinfo{issn}{0550-3213},
  \urlprefix\url{http://www.sciencedirect.com/science/article/pii/055032139190407O}.

\bibitem[{\citenamefont{Jain}(1989{\natexlab{b}})}]{Jain89b}
\bibinfo{author}{\bibfnamefont{J.~K.} \bibnamefont{Jain}},
  \bibinfo{journal}{Phys. Rev. B} \textbf{\bibinfo{volume}{40}},
  \bibinfo{pages}{8079} (\bibinfo{year}{1989}{\natexlab{b}}),
  \urlprefix\url{http://link.aps.org/doi/10.1103/PhysRevB.40.8079}.

\bibitem[{\citenamefont{Wu et~al.}(2017)\citenamefont{Wu, Shi, and
  Jain}}]{Wu16}
\bibinfo{author}{\bibfnamefont{Y.}~\bibnamefont{Wu}},
  \bibinfo{author}{\bibfnamefont{T.}~\bibnamefont{Shi}}, \bibnamefont{and}
  \bibinfo{author}{\bibfnamefont{J.~K.} \bibnamefont{Jain}},
  \bibinfo{journal}{Nano Letters} \textbf{\bibinfo{volume}{17}},
  \bibinfo{pages}{4643} (\bibinfo{year}{2017}), \bibinfo{note}{pMID: 28649831},
  \eprint{http://dx.doi.org/10.1021/acs.nanolett.7b01080},
  \urlprefix\url{http://dx.doi.org/10.1021/acs.nanolett.7b01080}.

\bibitem[{\citenamefont{Bandyopadhyay et~al.}(2018)\citenamefont{Bandyopadhyay,
  Chen, Ahari, Ortiz, Nussinov, and Seidel}}]{Bandyopadhyay18}
\bibinfo{author}{\bibfnamefont{S.}~\bibnamefont{Bandyopadhyay}},
  \bibinfo{author}{\bibfnamefont{L.}~\bibnamefont{Chen}},
  \bibinfo{author}{\bibfnamefont{M.~T.} \bibnamefont{Ahari}},
  \bibinfo{author}{\bibfnamefont{G.}~\bibnamefont{Ortiz}},
  \bibinfo{author}{\bibfnamefont{Z.}~\bibnamefont{Nussinov}}, \bibnamefont{and}
  \bibinfo{author}{\bibfnamefont{A.}~\bibnamefont{Seidel}},
  \bibinfo{journal}{Phys. Rev. B} \textbf{\bibinfo{volume}{98}},
  \bibinfo{pages}{161118} (\bibinfo{year}{2018}),
  \urlprefix\url{https://link.aps.org/doi/10.1103/PhysRevB.98.161118}.

\bibitem[{\citenamefont{Kim et~al.}(2019)\citenamefont{Kim, Balram, Taniguchi,
  Watanabe, Jain, and Smet}}]{Kim18}
\bibinfo{author}{\bibfnamefont{Y.}~\bibnamefont{Kim}},
  \bibinfo{author}{\bibfnamefont{A.~C.} \bibnamefont{Balram}},
  \bibinfo{author}{\bibfnamefont{T.}~\bibnamefont{Taniguchi}},
  \bibinfo{author}{\bibfnamefont{K.}~\bibnamefont{Watanabe}},
  \bibinfo{author}{\bibfnamefont{J.~K.} \bibnamefont{Jain}}, \bibnamefont{and}
  \bibinfo{author}{\bibfnamefont{J.~H.} \bibnamefont{Smet}},
  \bibinfo{journal}{Nature Physics} \textbf{\bibinfo{volume}{15}},
  \bibinfo{pages}{154} (\bibinfo{year}{2019}), ISSN \bibinfo{issn}{1745-2481},
  \urlprefix\url{https://doi.org/10.1038/s41567-018-0355-x}.

\bibitem[{\citenamefont{Balram and Jain}(2016)}]{Balram16b}
\bibinfo{author}{\bibfnamefont{A.~C.} \bibnamefont{Balram}} \bibnamefont{and}
  \bibinfo{author}{\bibfnamefont{J.~K.} \bibnamefont{Jain}},
  \bibinfo{journal}{Phys. Rev. B} \textbf{\bibinfo{volume}{93}},
  \bibinfo{pages}{235152} (\bibinfo{year}{2016}),
  \urlprefix\url{http://link.aps.org/doi/10.1103/PhysRevB.93.235152}.

\bibitem[{\citenamefont{Levin et~al.}(2007)\citenamefont{Levin, Halperin, and
  Rosenow}}]{Levin07}
\bibinfo{author}{\bibfnamefont{M.}~\bibnamefont{Levin}},
  \bibinfo{author}{\bibfnamefont{B.~I.} \bibnamefont{Halperin}},
  \bibnamefont{and} \bibinfo{author}{\bibfnamefont{B.}~\bibnamefont{Rosenow}},
  \bibinfo{journal}{Phys. Rev. Lett.} \textbf{\bibinfo{volume}{99}},
  \bibinfo{pages}{236806} (\bibinfo{year}{2007}),
  \urlprefix\url{http://link.aps.org/doi/10.1103/PhysRevLett.99.236806}.

\bibitem[{\citenamefont{Lee et~al.}(2007)\citenamefont{Lee, Ryu, Nayak, and
  Fisher}}]{Lee07}
\bibinfo{author}{\bibfnamefont{S.-S.} \bibnamefont{Lee}},
  \bibinfo{author}{\bibfnamefont{S.}~\bibnamefont{Ryu}},
  \bibinfo{author}{\bibfnamefont{C.}~\bibnamefont{Nayak}}, \bibnamefont{and}
  \bibinfo{author}{\bibfnamefont{M.~P.~A.} \bibnamefont{Fisher}},
  \bibinfo{journal}{Phys. Rev. Lett.} \textbf{\bibinfo{volume}{99}},
  \bibinfo{pages}{236807} (\bibinfo{year}{2007}),
  \urlprefix\url{http://link.aps.org/doi/10.1103/PhysRevLett.99.236807}.

\bibitem[{\citenamefont{Balram et~al.}(2018)\citenamefont{Balram, Barkeshli,
  and Rudner}}]{Balram18}
\bibinfo{author}{\bibfnamefont{A.~C.} \bibnamefont{Balram}},
  \bibinfo{author}{\bibfnamefont{M.}~\bibnamefont{Barkeshli}},
  \bibnamefont{and} \bibinfo{author}{\bibfnamefont{M.~S.}
  \bibnamefont{Rudner}}, \bibinfo{journal}{Phys. Rev. B}
  \textbf{\bibinfo{volume}{98}}, \bibinfo{pages}{035127}
  (\bibinfo{year}{2018}),
  \urlprefix\url{https://link.aps.org/doi/10.1103/PhysRevB.98.035127}.

\bibitem[{\citenamefont{Haldane}(1983)}]{Haldane83}
\bibinfo{author}{\bibfnamefont{F.~D.~M.} \bibnamefont{Haldane}},
  \bibinfo{journal}{Phys. Rev. Lett.} \textbf{\bibinfo{volume}{51}},
  \bibinfo{pages}{605} (\bibinfo{year}{1983}),
  \urlprefix\url{http://link.aps.org/doi/10.1103/PhysRevLett.51.605}.

\bibitem[{\citenamefont{Morf et~al.}(1986)\citenamefont{Morf, d'Ambrumenil, and
  Halperin}}]{Morf86}
\bibinfo{author}{\bibfnamefont{R.}~\bibnamefont{Morf}},
  \bibinfo{author}{\bibfnamefont{N.}~\bibnamefont{d'Ambrumenil}},
  \bibnamefont{and} \bibinfo{author}{\bibfnamefont{B.~I.}
  \bibnamefont{Halperin}}, \bibinfo{journal}{Phys. Rev. B}
  \textbf{\bibinfo{volume}{34}}, \bibinfo{pages}{3037} (\bibinfo{year}{1986}),
  \urlprefix\url{http://link.aps.org/doi/10.1103/PhysRevB.34.3037}.

\bibitem[{\citenamefont{Faugno et~al.}(2019)\citenamefont{Faugno, Balram,
  Barkeshli, and Jain}}]{Faugno19}
\bibinfo{author}{\bibfnamefont{W.~N.} \bibnamefont{Faugno}},
  \bibinfo{author}{\bibfnamefont{A.~C.} \bibnamefont{Balram}},
  \bibinfo{author}{\bibfnamefont{M.}~\bibnamefont{Barkeshli}},
  \bibnamefont{and} \bibinfo{author}{\bibfnamefont{J.~K.} \bibnamefont{Jain}},
  \bibinfo{journal}{Phys. Rev. Lett.} \textbf{\bibinfo{volume}{123}},
  \bibinfo{pages}{016802} (\bibinfo{year}{2019}),
  \urlprefix\url{https://link.aps.org/doi/10.1103/PhysRevLett.123.016802}.

\bibitem[{\citenamefont{Bonesteel et~al.}(1996)\citenamefont{Bonesteel,
  McDonald, and Nayak}}]{Bonesteel96}
\bibinfo{author}{\bibfnamefont{N.~E.} \bibnamefont{Bonesteel}},
  \bibinfo{author}{\bibfnamefont{I.~A.} \bibnamefont{McDonald}},
  \bibnamefont{and} \bibinfo{author}{\bibfnamefont{C.}~\bibnamefont{Nayak}},
  \bibinfo{journal}{Phys. Rev. Lett.} \textbf{\bibinfo{volume}{77}},
  \bibinfo{pages}{3009} (\bibinfo{year}{1996}),
  \urlprefix\url{https://link.aps.org/doi/10.1103/PhysRevLett.77.3009}.

\bibitem[{\citenamefont{McDonald and Haldane}(1996)}]{McDonald96}
\bibinfo{author}{\bibfnamefont{I.~A.} \bibnamefont{McDonald}} \bibnamefont{and}
  \bibinfo{author}{\bibfnamefont{F.~D.~M.} \bibnamefont{Haldane}},
  \bibinfo{journal}{Phys. Rev. B} \textbf{\bibinfo{volume}{53}},
  \bibinfo{pages}{15845} (\bibinfo{year}{1996}),
  \urlprefix\url{https://link.aps.org/doi/10.1103/PhysRevB.53.15845}.

\bibitem[{\citenamefont{Balram}(2016)}]{Balram16c}
\bibinfo{author}{\bibfnamefont{A.~C.} \bibnamefont{Balram}},
  \bibinfo{journal}{Phys. Rev. B} \textbf{\bibinfo{volume}{94}},
  \bibinfo{pages}{165303} (\bibinfo{year}{2016}),
  \urlprefix\url{http://link.aps.org/doi/10.1103/PhysRevB.94.165303}.

\bibitem[{\citenamefont{T\ifmmode~\mbox{\H{o}}\else \H{o}\fi{}ke and
  Jain}(2007)}]{Toke07}
\bibinfo{author}{\bibfnamefont{C.}~\bibnamefont{T\ifmmode~\mbox{\H{o}}\else
  \H{o}\fi{}ke}} \bibnamefont{and} \bibinfo{author}{\bibfnamefont{J.~K.}
  \bibnamefont{Jain}}, \bibinfo{journal}{Phys. Rev. B}
  \textbf{\bibinfo{volume}{75}}, \bibinfo{pages}{245440}
  (\bibinfo{year}{2007}),
  \urlprefix\url{http://link.aps.org/doi/10.1103/PhysRevB.75.245440}.

\bibitem[{\citenamefont{Wu et~al.}(2015)\citenamefont{Wu, Sodemann, MacDonald,
  and Jolicoeur}}]{Wu15}
\bibinfo{author}{\bibfnamefont{F.}~\bibnamefont{Wu}},
  \bibinfo{author}{\bibfnamefont{I.}~\bibnamefont{Sodemann}},
  \bibinfo{author}{\bibfnamefont{A.~H.} \bibnamefont{MacDonald}},
  \bibnamefont{and}
  \bibinfo{author}{\bibfnamefont{T.}~\bibnamefont{Jolicoeur}},
  \bibinfo{journal}{Phys. Rev. Lett.} \textbf{\bibinfo{volume}{115}},
  \bibinfo{pages}{166805} (\bibinfo{year}{2015}),
  \urlprefix\url{http://link.aps.org/doi/10.1103/PhysRevLett.115.166805}.

\end{thebibliography}
\bibliographystyle{apsrev}

\end{document}